\documentclass{article}
\newcommand{\be}{\begin{equation}}
\newcommand{\ee}{\end{equation}}

\def\bea{\begin{eqnarray}}
\def\eea{\end{eqnarray}}
\def\bean{\begin{eqnarray*}}
\def\eean{\end{eqnarray*}}
\def\jrn#1#2#3#4{{#1} {\bf #2} (#4) #3}
\def\PRL{\it Phys. Rev. Lett.}
\def\PLB{\it Phys. Lett. B}
\def\PRD{\it Phys. Rev. D}
\def\NPB{\it Nucl. Phys. B}
\def\addressfont{\small\itshape{}}
\def\address#1{\expandafter\def\expandafter\@aabuffer\expandafter
	{\@aabuffer{\addressfont{#1\par}\par}\relax\par
	\vspace*{13pt}}}
\def\abstracts#1{
\begin{center}
{\begin{minipage}{4.2truein}
                 \footnotesize
                 \parindent=0pt #1\par
                 \end{minipage}}\end{center}
                 \vskip 2em \par}

\begin{document}

\title{Neutrinos: Windows to  Planck Physics\footnote{\uppercase{T}his work  was supported in part
by the US Department of Energy under grant DE-FG02-97ER41029}}

\author{P. Ramond}

\address{Institute for Fundamental Theory, \\
Physics Department, University of Florida \\ 
P.O. Box 118440, Gainesville, FL 32611, USA\\ 
E-mail: ramond@phys.ufl.edu}

\maketitle

\abstracts{
After recalling some elegant contributions of the late Freydoon Mansouri, we turn to neutrino physics and use a {\it modicum} of grand unification to relate quark and lepton mixing matrices. We advocate an expansion for the MNS matrix, {\it \`a la}  Wolfenstein, and  argue that in a wide class of models, $\theta_{13}$ is a Cabibbo mixing effect. Also the large neutrino mixing angles reflect the mass patterns of the right-handed neutrinos near the Planck scale, and provide evidence for family structure at that scale.}

\section{A Few words about Freyd}
In my mind, one  word captures Freyd Mansouri: elegance in physics and in life.  I will mention two very beautiful papers and a recent one which make the point. The first is 
\vskip .2cm
$\bullet$ ``{\it Dynamics Underlying Duality and Gauge Invariance in the Dual Resonance Model}"  with L. N. Chang, Dec 1971.

\noindent In that paper, the authors use the conformal invariance of the Nambu-Goto action to linearize it, and obtain the well-known constraints. This was written at a time where most were unaware of the conformal symmetry of two-dimensional systems. These constraints were later explicitly solved in  the  classic  GGRT paper.  The second paper is 
\vskip .2cm
$\bullet$ ``{\it Unified Geometric Gravity and Supergravity}" with S. MacDowell, Feb 1977,
where a beautiful geometric framework is introduced to describe both gravity and supergravity.  These papers did not get the publicity they deserved as Freyd always found it distasteful to promote his own work, a virtue not shared by most of his contemporaries. Finally I mention a more recent paper

$\bullet$ ``{\it The Cosmological Constant Problem, The Spontaneously Broken Symmetry, And The Generalized Rest Energy}~", Dec 2001, 
where he derives the generalization of Einstein's famous formula in DeSitter space:

{$$m^2c^4~=~E^{}_0(E^{}_0-3 ~{E^{}_\Lambda})+ ~{E^2_\Lambda(2+s-s^2_{})}$$}
Then he goes on to say: 
``{\it For particles of small mass, such as neutrinos, assuming that they are massive, such a deviation from the standard rest energy will significantly affect their kinematics at very high energies. It may be possible to test this proposal in not too distant a future}". This brings me to the subject of this talk, neutrino physics.

\section{Neutrino data}
The results\cite{FOGLI} of five years of experiments\cite{SKatm,SKsol,SNO,CHOOZ,kamland,K2K} on neutrinos can be summarized by the MNS neutrino mixing matrix 

$$\mathcal{ U}^{}_{MNS}~=~\pmatrix{\cos\phi&\sin\phi&\epsilon \cr
-\cos\theta~\sin\phi&\cos\theta~\cos\phi&\sin\theta \cr
\sin\theta~\sin\phi&-\sin\theta~\cos\phi&\cos\theta}\ ,$$
in which 

$$ \sin^2 2\theta~>~0.85\ ;\qquad  0.30~<~ \tan^2\phi~< ~0.65\ ;\qquad \vert~\epsilon~\vert^2_{}~<~0.005\ .$$
The most striking feature of this mixing matrix is the appearance of two large mixing angles, one of which may well be maximal, and 
only one small mixing angle. That angle has not yet been measured, its limiting value being set by the CHOOZ\cite{CHOOZ} reactor experiment. It is a strange situation where a matrix is diagonalized by two large and one small angle: it may hint (unlike the quark sector) at a non-Abelian structure in the mixing pattern. The following presents a framework in which to study such possibilities.

Experiments have also set limits on the mass values of the neutrinos.  WMAP\cite{WMAP} is the only experiment to set a limit on the absolute value of the neutrino masses

$$\sum_i~m^{}_{\nu_i}~<~.71~{\rm eV}$$
The others, which  detect  neutrino oscillations, are only sensitive to mass differences with the results

$$\Delta m^2_\odot~=~\vert~m^2_{\nu_1}-m^2_{\nu_2}~\vert~\sim~7.\times 10^{-5}_{}~{\rm eV^2}$$
$$\Delta m^2_\oplus~=~\vert~m^2_{\nu_2}-m^2_{\nu_3}~\vert~\sim~3.\times 10^{-3}_{}~{\rm eV^2}$$
Since we have no absolute mass measurements, the present data are compatible with three possible mass patterns\cite{SMIRNOV}:

\vskip .3cm
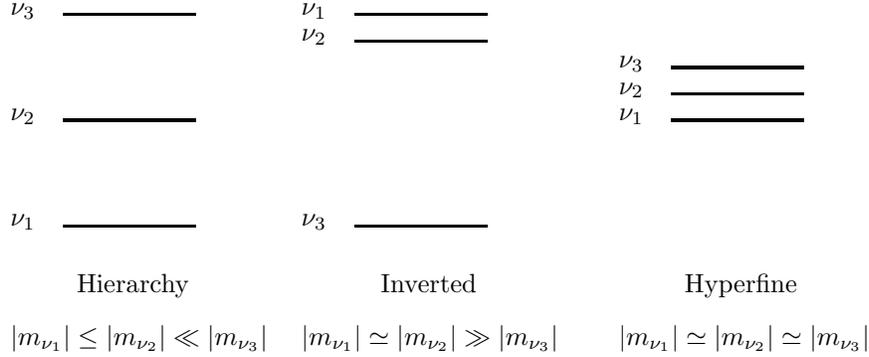
\begin{picture}(200,170)

\thicklines
\put(20,70){\line(1,0){50}}
\put(20,110){\line(1,0){50}}
\put(20,150){\line(1,0){50}}
\put(0,70){\shortstack{$\nu_1$}}
\put(0,110){\shortstack{$\nu_2$}}
\put(0,150){\shortstack{$\nu_3$}}
\put(25,45){\shortstack{Hierarchy}}
\put(0,25){\shortstack{$|m_{\nu _1}| 
\leq |m_{\nu _2}| \ll |m_{\nu _3}|$}}

\put(130,70){\line(1,0){50}}
\put(130,140){\line(1,0){50}}
\put(130,150){\line(1,0){50}}
\put(110,70){\shortstack{$\nu_3$}}
\put(110,140){\shortstack{$\nu_2$}}
\put(110,150){\shortstack{$\nu_1$}}
\put(140,45){\shortstack{Inverted}}
\put(110,25){\shortstack{$|m_{\nu _1}| 
\simeq |m_{\nu _2}| \gg |m_{\nu _3}|$}}

\put(250,110){\line(1,0){50}}
\put(250,120){\line(1,0){50}}
\put(250,130){\line(1,0){50}}
\put(230,110){\shortstack{$\nu_1$}}
\put(230,120){\shortstack{$\nu_2$}}
\put(230,130){\shortstack{$\nu_3$}}
\put(255,45){\shortstack{Hyperfine}}
\put(230,25){\shortstack{$|m_{\nu _1}| 
\simeq |m_{\nu _2}| \simeq |m_{\nu _3}|$}}
\end{picture}
\vskip .2cm

\noindent To determine which of these patterns Nature chooses will require the observation of neutrinoless double $\beta$ decay. 

 \section{Mixings in the Standard Model}
In the Standard Model, masses and mixings of the quarks and charged leptons are the consequence of the spontaneous breakdown of electroweak symmetry,  $\Delta I_{\rm w}=1/2$. This yields the Quark Yukawa Matrices

$$
{\mathcal M}^{(2/3)}_{}~={\mathcal U}^{}_{2/3}\,
\pmatrix{m^{}_u&0&0\cr 0&m^{}_c&0\cr 0&0&m_t^{}}
\,{\mathcal V}^{\dagger}_{2/3}\ ,$$
$$
{\mathcal M}^{(-1/3)}_{}~=~{\mathcal U}^{}_{-1/3}\,
\pmatrix{m^{}_d&0&0\cr 0&m^{}_s&0\cr 0&0&m_b^{}}
\,{\mathcal V}^{\dagger}_{-1/3}\ ,
$$
from which the observable CKM quark mixing matrix is obtained
 $$
{\mathcal U}^{}_{CKM}~=~{\mathcal U}^{\dagger}_{2/3}\,{\mathcal U}^{}_{-1/3}\ .
$$
Experimentally, it is nearly equal to the unit matrix, up to small powers of $\lambda$, the Cabibbo angle.  
Family mixing treats  quarks of charge $2/3$ and $-1/3$ almost alike. This is exploited in the Wolfenstein\cite{WOLF} expansion which uses the unit matrix as a starting point. 

The same electroweak breaking produces the  $\Delta I_{\rm w}=1/2$  charged Leptons Yukawa Matrix
$${\mathcal M}^{(-1)}_{}~=~{\mathcal U}^{}_{-1}\,
\pmatrix{m^{}_e&0&0\cr 0&m^{}_\mu&0\cr 0&0&m_\tau^{}}
\,{\mathcal V}^{\dagger}_{-1}\ .$$
In the original formulation of the Standard Model, these  mixing matrices are unobservable and the neutrinos are massless. However by 
 adding one right-handed neutrino per family, as suggested by Pati-Salam unification\cite{PS}, we generate the $\Delta I_{\rm w}=1/2$  neutral leptons Yukawa matrix

 $${\mathcal M}^{(0)}_{ ~{Dirac}}~=~{\mathcal U}^{}_{0}\,{\mathcal D}_0^{}\,
{\mathcal V}^{\dagger}_{0}~=~{\mathcal U}^{}_{0}\,
\pmatrix{m^{}_1&0&0\cr 0&m^{}_2&0\cr 0&0&m_3^{}}\,{\mathcal V}^{\dagger}_{0}\ .
$$
In the absence of any other masses, this produces the Maki-Nakagawa-Sakata (MNS)  matrix, the equivalent of the 
CKM matrix in the lepton sector

$${\mathcal U}^{}_{MNS}~=~ {\mathcal U}^{\dagger}_{-1}\,{\mathcal U}^{}_{0}\ ,$$
of the same form as in the quark sector. However, unlike in the quark sectors, the right-handed neutrinos can have a $\Delta I_{\rm w}^{}=0$ Majorana mass, unprotected by the gauged electroweak symmetry, although it breaks the global lepton number symmetry

$${\mathcal M}^{(0)}_{ ~{Majorana}}~\sim~ \Delta I_{\rm w}^{}=0\ .$$
If  the Standard Model has a natural cut-off  $\Lambda$, we expect those masses to be of that  order. If $\Lambda$  is much larger than the scale of electroweak breaking,  the seesaw mechanism\cite{SEESAW}  naturally suppresses the neutrino masses over that of their charged partners by the ratio of these scales:

\bea{\mathcal M}^{(0)}_{ ~{Seesaw}}&=&{\mathcal U}^{}_{0}\,
{\mathcal D}_0^{}\,{\mathcal V}^{\dagger}_{0}\,\frac{1}{{\mathcal M}^{(0)}_{ ~{Majorana}}}\,
{\mathcal V}^{*}_{0}\,{\mathcal D}_0^{}\,{\mathcal U}^{T}_{0}\cr
&~&\cr
&\equiv&{\mathcal U}^{}_{0}\,
 ~{{\mathcal C}}\,\,{\mathcal U}^{T}_{0}\ ,\nonumber\eea
which  defines the {\em Central Matrix}\cite{DLR}  $\mathcal C$, whose properties contain the added subtelties of the seesaw.  
  It is diagonalized by a unitary matrix $\mathcal F$ 
 
$$ 
{\mathcal C}~= ~{\mathcal F}\,{\mathcal D}^{}_\nu\, ~{\mathcal F}^{\,T}_{}\ ,
$$
 its eigenvalues being the physical neutrino masses  

$$
{\mathcal D}_\nu^{}~=~
\pmatrix{m^{}_{\nu_1}&0&0\cr 0&m^{}_{\nu_2}&0\cr 0&0&m_{\nu_3}^{}}\ .
$$
The MNS  matrix now becomes 

$${\mathcal U}^{}_{MNS}~=~ {\mathcal U}^{\dagger}_{-1}\,{\mathcal U}^{}_{0}\, ~{\mathcal F}\ ,$$
 where  $ ~{\mathcal F}$ is the new $\Delta I_{\rm w}=0$ ingredient  that comes from the seesaw mechanism. 

If the matrices  from the $\Delta I_{\rm w}=1/2$ sector contain only small angles, as the CKM matrix and mass hierarchies might suggest, it is natural to ask  if  the large angles come from $~{\mathcal F}$.  Furthermore, as the data indicates  one small and two large angles, we may also ask if this arrangement is natural in a $3\times 3$ matrix. Indeed, if mixing matrices are generated by textures of zeros\cite{RRR}, then such an arrangement is difficult to imagine as it would produce either three, two or no small angles. Any other arrangement  would point to a symmetry beyond texture zeros\cite{KING}. It is therefore compelling to investigate the origin of the one small angle. 

From the texture point of view, one large mixing angle is generic for a $(3\times 3)$ matrix, and  in a class of models, 
it is natural for  $ {\mathcal F}$ to contain only one large angle, with the second large angle hiding in the diagonalization of the {\it right-handed} quarks and charged leptons. 

\section{Hints from Grand Unification}
To link  the MNS and CKM matrices, we need to  use some notions of Grand Unification. In its simplest form, it is well-known that the Standard Model families form $SU(5)$\cite{SU5} and $SO(10)$\cite{SO10} multiplets, the latter containing one right-handed neutrino per family (going up the exceptional group series, one arrives at $E_6$\cite{E6} with three right-handed neutrinos per family). The simplest  Higgs structure of these models relates quark and lepton $\Delta I^{}_{\rm w}=1/2$ Yukawa matrices:

 $$ SU(5)~~~:~~~~~~~~{\mathcal M}^{(-1/3)}_{}~\sim~{\mathcal M}^{(-1)\,T}_{}\ .$$
 $$ SO(10):~~~~~~~~{\mathcal M}^{(2/3)}_{}~\sim~{\mathcal M}^{(0)}_{ ~{Dirac}}\ . $$

 Applying these relations to the Yukawa matrices yields  

$${\mathcal U}^{}_{-1/3}~\sim~{\mathcal V}^{*}_{-1}\ ;\qquad  {\mathcal U}^{}_{2/3}~\sim~{\mathcal U}^{}_{0}\ .$$
It follows that 

\bea\nonumber{\mathcal U}^{}_{MNS}&=& {\mathcal U}^{\dagger}_{-1}\,{\mathcal U}^{}_{0}\,{\mathcal F}~=~{\mathcal U}^{\dagger}_{-1}\,{\mathcal U}^{}_{-1/3}\,{\mathcal U}^{\dagger}_{CKM}\,{\mathcal F}\\
\nonumber&=& {\mathcal V}^T_{-1/3}\,{\mathcal U}^{}_{-1/3}\,{\mathcal U}^{\dagger}_{CKM}\,{\mathcal F}\ .\eea
This is the desired relation between the quark and lepton mixing matrices. They are seen to differ in two ways, first by the seesaw which requires $\mathcal F$ matrix, and also by the right-handed matrix that diagonalizes the charge $-1/3$ quark sector.  

Accordingly we group models of Yukawa couplings in two categories, those for which the charge $-1/3$ Yukawa couplings 
are symmetric, and those for which they aren't. 

 \begin{itemize}
\item In models where ${\mathcal M}^{}_{-1/3}$ is symmetric\cite{KING},  we have 

$${\mathcal U}^{}_{-1/3}~=~{\mathcal V}^*_{-1/3}\ ,$$ 
and  the two matrices are simply related

$${\mathcal U}^{}_{MNS}~=~ {\mathcal U}^{\dagger}_{CKM}\,\,{\mathcal F}\ .$$
In this case, $\mathcal F$ must contain two large mixing angles, and therefore be very special,  implying perhaps a non-Abelian  structure.

\item In the other case, one or two large angles can hide in ${\mathcal V}_{-1/3}$. This occurs in models where assymmetric Yukawas are natural. We mention two possibilities. The first is that of {\em family cloning}\cite{BR}  where there is one standard model gauge group for each family. It is broken down to one gauge group by a  tri-chiral order parameter\cite{LR} which naturally yields one diagonal color group 
$SU(3)_1\times SU(3)_2\times SU(3)_3~\rightarrow~SU(3)_{1+2+3}$
 but can only break  the weak $SU(2)$ to two groups $SU(2)_1\times SU(2)_2\times SU(2)_3~\rightarrow~SU(2)_{1+2}\times SU(2)_3$, where the subscript denotes the family number. This results, at some intermediate scale, into the following gauge 
symmetry: 
$$SU(2)_{1+2}\times SU(2)_3\times SU(3)_{1+2+3}\ ,$$
which naturally produces  asymmetric Yukawa matrices and one large angle in $U_{-1}$. 

Another, more generic class of model in which this appears, are  of the Froggatt-Nielsen\cite{FN} type where Cabibbo suppression is related to the dimensions of the Yukawa operators.  By expressing  the charge $-1/3$ quark mass ratios as powers of the Cabibbo angle, 

$$\frac{m_s}{m_b}~\sim~\lambda^2_{}\ ,\qquad \frac{m_d}{m_b}~\sim~\lambda^4_{}\ ,$$
and assuming similar Cabibbo suppression in both $U_{2/3}$ and $U_{-1/3}$,  the suggestive order of magnitude pattern appears 

$${\mathcal M}^{(-1/3)}_{}~=~\pmatrix{~{\lambda^4_{}}& ~{\lambda^3_{}}& ~{\lambda^3_{}}\cr
 ~{\lambda^3_{}}& ~{\lambda^2_{}}& ~{\lambda^2_{}}\cr
 ~{\lambda}&1&1}\ .$$
The {\small 32}  matrix element is not suppressed compared to the {\small 33},  so that diagonalization requires a large angle {\em in the right-handed sector}, naturally predicting\cite{ILR}   unsuppressed mixing between mu- and tau-type neutrinos,  although not explaining its near maximality. In this model,  before Cabibbo corrections, 

$${\mathcal U}^{}_{MNS}~=~\pmatrix{1&0&0\cr 0&\cos\theta&\sin\theta\cr 0&-
\sin\theta&\cos\theta} \,\,{\mathcal F}\ ,$$
and $\mathcal F$ need  contain only one large angle, that responsible for the solar neutrino deficit. 
While  not maximal, it is close to $30^o$. One amusing possibility is to posit a  Central matrix  of the form 

$${\mathcal C}~=~\pmatrix{0&1&0\cr 1&1&0\cr 0&
0&1}\ ,$$
which is diagonalized by the golden ratio angle\cite{DLR} 

$$\tan\eta~=~\frac{1-\sqrt{5}}{1+\sqrt{5}}\ ,$$
whose pleasing nature has been known to architects for millenia. Could it also be pleasing to {\em the} Architect?
\end{itemize}
In either case, we may insert the two large angles in the form 

$${\mathcal U}^{}_{MNS}~=~\pmatrix{1&0&0\cr 0&\frac{1}{\sqrt{2}}&-\frac{1}{\sqrt{2}}\cr 0&\frac{1}{\sqrt{2}}&\frac{1}{\sqrt{2}}}\pmatrix{\cos\eta&\sin\eta&0\cr -\sin\eta&\cos\eta&0\cr0&0&0} \ .$$
In models with symmetric charge $-1/3$ Yukawas, the Cabibbo effects  from the CKM matrix come by multiplication from the left, yielding the order of magnitude of  the CHOOZ angle  

$$\theta^{}_{13}~\sim~ \frac{\lambda}{\sqrt{2}}\ ,$$
which is satisfying near the experimental limit. 

In models where the atmospheric angle hides in the right-handed sector, the CKM matrix is sandwiched between these two matrices; it follows that the CHOOZ angle is determined by the Cabibbo structure of the right-handed matrix ${\mathcal V}_{-1/3}$. The CKM matrix gives only a very small contribution of order $\lambda^3$ to $\theta^{}_{13}$, but   

$$\theta^{}_{13}~\sim~ \frac{\lambda^p}{\sqrt{2}}\ ,$$
where $\lambda^p$ is the order of magnitude of the {\small 21} matrix element of ${\mathcal V}_{-1/3}$. 
 In these models,  the value of that angle depends on the hithero unobservable matrix which diagonalizes the right-handed down quarks. 

Models with grand unified relations between the quark and lepton mixing matrices suggest a Wolfenstein parametrization of the MNS matrix, not using as a starting point  a unit matrix as for the CKM matrix, but a  matrix that contains the two large mixing angles. The CHOOZ angle will be determined as some power of the Cabibbo angle.  Our favorite starting point is thus

$${\mathcal U}^{}_{MNS}~=~\pmatrix{\cos\eta&\sin\eta&0\cr -\frac{\sin\eta}{\sqrt{2}}&\frac{\cos\eta}{\sqrt{2}}&-\frac{1}{\sqrt{2}}\cr -\frac{\sin\eta}{\sqrt{2}}&\frac{\cos\eta}{\sqrt{2}}&\frac{1}{\sqrt{2}}} ~+~\cdots\ ,$$
where the dots contain Cabibbo-like effects (we could have chosen $\eta=\pi/6$ as well for this purpose, but we prefer the golden mean for \ae sthetic reasons). 

\section{The origin of large angles}
As we have seen,  $\mathcal F$  contains at least  one large angle. Let us recall that it diagonalizes the Central matrix which is given by 

$$ ~{\mathcal C}~\equiv~ ~{{\mathcal D}_0^{}}\,{\mathcal V}^{\dagger}_{0}\,
\frac{1}{{\mathcal M}^{(0)}_{ ~{Majorana}}}\,{\mathcal V}^{*}_{0}\, ~{{\mathcal D}_0^{}}\ ;$$
since we believe the $\Delta I_{\rm w}=1/2$ neutral Dirac mass to be hierarchical, the large angle(s) in its diagonalization hints at some correlation between its Majorana and Dirac components. Let us see how this works in a simplified $(2\times 2)$ case. We start with  hierarchical Dirac eigenvalues\cite{DLR,SMIR}, 

 $$
{\mathcal D}^{}_0~=~m\pmatrix{a\, ~{\lambda^\alpha_{}}&0\cr 0&1}\ ,$$
 from which it follows that 

  
$$
{\mathcal C}~=~\pmatrix{(\frac{c^2}{M_1}+\frac{s^2}{M_2})\,a^2_{}\,
{\lambda^{2\alpha}_{}}&(\frac{c\,s}{M_1}-\frac{c\,s}{M_2})\,a\,{\lambda^\alpha_{}}\cr
(\frac{c\,s}{M_1}-\frac{c\,s}{M_2})\,a\,{\lambda^\alpha_{}}&
(\frac{s^2}{M_1}+\frac{c^2}{M_2})}\ ,
$$
where $M_{1,2}$ are the eigenvalues of the Majorana matrix, and $c,s$ stand for the sine and cosine of some angle (not the one that diagonalizes the matrix).  It is diagonalized by a large mixing angle in two different cases:
 \begin{itemize}
\item All its matrix elements are of the same order of magnitude, ${\mathcal C}_{11} \sim{\mathcal C}_{22}\sim{\mathcal C}_{12} $. This can happen as long as  
 
$$ s~\sim~b\, ~{\lambda^\alpha_{}}\ ,\qquad c~\sim~1\ ,\qquad 
\frac{M_1}{M_2}~\sim~ ~{\lambda^{2\beta}_{}}\ ,$$
implying a  correlated hierarchy between the $\Delta I_{\rm w}=1/2$ and $\Delta I_{\rm w}=0$ sectors.


 \item Its diagonal matrix elements are much smaller than its off-diagonal ones (level-crossing case) $
{\mathcal C}_{11}\ , {\mathcal C}_{22} \prec {\mathcal C}_{12}$. It is possible if 

$$
\lambda^\alpha ~\prec~ s ~\preceq~ \lambda^{\alpha-\beta}\ ,
$$
 yielding a Central matrix  like

$$\frac{\lambda^\alpha~m^2}{\sqrt{-M_1 M_2}}\,
\pmatrix{0&a\cr a&0}\ ,$$
which displays maximal mixing.  A way to arrange is is to take the two right-handed neutrinos to be Dirac partners. 

\end{itemize}
In the first case, we are faced with a large  hierarchy of the $\Delta I_{\rm w}=0$ masses. If $\mathcal F$ contains two large angles, all three right-handed neutrinos  show hierarchies. It follows that the hierarchy is present even at the Standard Model cut-off. As a welcome by-product, we note that  leptogenesis  requires right-handed neutrinos with mass of the order of $10^9$ GeV, much less than $\Lambda$, implying a strong mass hierarchy .  

Much remains to be done, but it is clear that searching for the origin of the large mixing angles brings us to consider GUT-scale physics as it relates to the masses of the right-handed neutrinos. 

The most exciting aspect of neutrino physics is that it brings to the Standard Model scales near the natural scale of gravity. To include gravity at the quantum level, we need superstrings which carry the message of extra symmetries associated with extra dimensions and 
supersymmetry between fermions and bosons.  
Since the right-handed neutrinos are a bridge between Planck and electroweak scale physics, model builders should forget about the Standard Model, and first incorporate  right-handed neutrinos devoid of any electroweak quantum numbers into their favorite braneworld or extra dimension scenario. Having done so, electroweak physics will be added by completing the $SO(10)$ multiplets\cite{AB,SOTEN}. 
 
In these exciting times with a flow of new experimental information in particle physics, let us hope that neutrinos will prove the key to unravelling  the long-standing riddle of flavor physics. 

\section{Acknowledgements} I wish to express my condoleances to Freyd's family, as well as my thanks to Professor Wijewardhana and his colleagues for organizing such an excellent conference under difficult circumstances. I also wish to thank A. Datta, L. Everett  and F-S Ling for useful discussions.

\end{document}